\documentclass[%
superscriptaddress,
preprint,
showpacs,preprintnumbers,
 amsmath,amssymb,
prb,
]{revtex4-1}

\usepackage{graphicx}
\usepackage{color}
\usepackage{dcolumn}
\usepackage{bm}
\usepackage{hyperref}

\begin{document}

\title{Electronic and Transport Property of Phosphorene Nanoribbon}

\author{Qingyun Wu}
\affiliation{Department of Physics, National University of Singapore, Singapore 117542, Singapore}%
\author{Lei Shen}%
\email{shenlei@nus.edu.sg}
\affiliation{Engineering Science Programme, Faculty of Engineering, National University of Singapore, Singapore 117579, Singapore}%
\author{Ming Yang}
\affiliation{Department of Physics, National University of Singapore, Singapore 117542, Singapore}%
\author{Yongqing Cai}
\affiliation {Institute of High Performance Computing, ASTAR, Singapore 138632, Singapore}%
\author{Zhigao Huang}
\affiliation{College of Physics and Energy, Fujian Normal University, Fuzhou 350007, People's Republic of China}%
\author{Yuan Ping Feng}
\email{phyfyp@nus.edu.sg}
\affiliation{Department of Physics, National University of Singapore, Singapore 117542, Singapore}%

\date{\today}

\begin{abstract}
By combining density functional theory and nonequilibrium Green's function, we study the electronic and transport properties of monolayer black phosphorus nanoribbons (PNRs). First, we investigate the band-gap of PNRs and its modulation by the ribbon width and an external transverse electric field. Our calculations indicate a giant Stark effect in PNRs, which can switch on transport channels of semiconducting PNRs under low bias, inducing an insulator-metal-transition. Next, we study the transport channels in PNRs via the calculations of the current density and local electron transmission pathway. In contrast to graphene and MoS$_{2}$ nanoribbons, the carrier transport channels under low bias are mainly located in the interior of both armchair and zigzag PNRs, and immune to a small amount of edge defects. Lastly, a device of the PNR-based dual-gate field-effect-transistor, with high on/off ratio of 10$^3$, is proposed based on the giant electric field tuning effect.
\end{abstract}

\pacs{73.61.Cw, 73.22.-f, 73.63.-b, 71.15.Mb}
\maketitle

\section{Introduction}
Although many two dimensional (2D) materials, such as graphene and MoS$_2$, have good carrier mobilities or high on/off ratio,\cite{Schwierz2010NN,Radisavljevic2011NN} a fundamental dilemma hampers their nanoribbon device applications: Whereas robust transport is needed to immune the edge-defect perturbation, modulation of transport by an electric field (or gate voltage) is desired for on/off transistors. One of the reasons is that the transport channels in most nanoribbons are located at two edges.\cite{Kim2008NN,Zheng2008PRB,Yue2012JPCM} A small amount of edge disorder or defects, such as vacancies and impurities, can strongly suppress the carrier mobility in the transport channels because of Coulomb blockage or scattering.\cite{Sols2007PRL,Shen2012PRB} Another reason is that many nanoribbons have a metallic edge-morphology, such as zigzag graphene \cite{Wang2008PRL,Son2006PRL} and zigzag MoS$_2$ nanoribbons \cite{Li2008JACS}, which has little response to the gate voltage, resulting in very low on/off ratio of graphene/MoS$_2$ field-effect-transistors (FETs). Actually, such different transport behaviors due to different edge-morphologies in nanoribbons are similar to the chirality of carbon nanotubes (CNTs), which hinders the development of CNT FETs till now. Very recently, layered black phosphorus (phosphorene) and its sisters \cite{Zhu2014PRL,Zhu2015PRB,Guan2014PRL2,Guan2014PRL} have attracted much attention because of its unique electronic properties \cite{Li2014NN,Liu2014AN,Qiao2014NC,Rodin2014PRL,Zhu2014PRL,Reich2014Nature,Churchill2014NN,Koenig2014APL,Fei2014NL,Fei2015PRB,Tran2014PRB,Tran2014PRB2,Guo2014JPCC,Peng2014PRB,Peng2014MRE,Wang2015PRB,Cai2014SR} and thermoelectronic properties\cite{Zhang2014SR,Cai2015AFM}. In bulk form, black phosphorus consists of puckered honeycomb layers of phosphorus atoms which are held together via van der Waals interactions, similar to graphite.\cite{brown1965} It is a direct band gap semiconductor with an energy gap of 0.3 eV,\cite{Akahama1983JPSJ} while the gap of monolayer phosphorene is 1.5 eV.\cite{Qiao2014NC} Field-effect-transistors (FETs) based on a few layers of phosphorene were found to have on/off ratio up to 10$^{5}$ [ref. 21] and carrier mobilities as high as 1,000 cm$^{2}$/Vs [ref. 15] at room temperature. It seems promising for phosphorene to compete with other hot contenders, graphene and layered MoS$_{2}$, for next generation semiconductor devices.

One-dimensional nanoribbons etched or patterned from their parent 2D materials offer additional tenability of their electronic properties through the quantum confinement effect .\cite{Dolui2012AN,Li2008JACS,Son2006PRL,Yue2012JPCM,Zhang2008PRB,Zheng2008PRB,Park2008Nl} However, as mentioned above, most nanoribbons are chiral and their transport properties are dominated by edge states. Taking MoS$_{2}$ nanoribbon as an example, it is known that armchair MoS$_{2}$ nanoribbons are semiconducting while zigzag MoS$_{2}$ nanoribbons are metallic.\cite{Li2008JACS,Dolui2012AN} The carrier mobility of armchair MoS$_{2}$ nanoribbons can be strongly suppressed by the scattering of edge defects .\cite{Dolui2012AN,Yue2012JPCM,Cai2014JACS} Due to the limit of the chirality and edge transport channels, one have to fabricate high quality \emph{armchair} MoS$_{2}$ nanoribbons in the experiment for making transistors with good performance.\cite{Radisavljevic2011NN} Clearly, it is very difficult to fabricate edge-defect-free nanoribbons with controlled chirality as well. Compared to graphene and 2D dichalcogenides, phosphorene shows certain advantages. Even though there have been many studies in electronic structures of phosphorene ribbons,\cite{Carvalho2014EPL,Li2014JPCC,Ramasubramaniam2014PRB,Han2014NL,Du2015SR,Maity2014arxiv,Peng2014JAP} further \emph{transport} studies compared to graphene and MoS$_{2}$ are less,\cite{Schwierz2010NN} but essential to fully understand its intriguing properties and find avenues to tune these properties for various device applications.

Inspired by ubiquitous gate-control in everyday semiconductor devices, in this work, we report the robust central transport in achirial PNRs under low bias, and its effective electric-field-modulation through a giant Stark effect. By using the giant electric field tuning effect, we propose a PNRs-based dual-gate FET, which is expected to have high on/off ratio. This article is organized as following:
The computational details are given in Sec. \textbf{II.} and Sec. \textbf{X.} We present details of the optimized geometric structures in Sec. \textbf{III.} The calculated energy band structures and band gap tuning by the ribbon width or electric field are investigated in Sec. \textbf{IV.} and \textbf{VI.} Transport properties, including in transport channels and their defect/bias effects, are studied in Sec. \textbf{V.} and subsections. The PNR-based electronic devices, in Sec. \textbf{VII.}, are proposed. Lastly, the conclusions and discussions of this article are shown in Sec. \textbf{VIII.}

\section{Computational details}
First-principles calculations based on density functional theory (DFT) and the non-equilibrium Green's function (NEGF) formalism were carried out to study electronic and transport properties of PNRs, which was implemented in the \textsc{Atomistix ToolKit} package.\cite{atk1,atk2} Geometry optimization was done until all atomic forces are smaller than 0.01 eV/{\AA}. The generalized gradient approximation (GGA) with Perdew-Burke-Ernzerhof (PBE) functional\cite{pbe} was used for the exchange-correlation functional. The electron wave function was expanded using a double-$\zeta$ polarized (DZP)basis set. A mesh cut-off of 150 Ry and a Monkhorst-Pack $k$-point grid\cite{Monkhorst} of $1\times1\times9$ were employed in the electronic calculations. During the transport calculations, a $k$-point grid of $1\times1\times100$ was adopted. Vacuum layers of 15 \AA~both in plane and out of plane of the ribbons were used to avoid the interaction between periodic images. Regarding the width of PNRs to N=35, band gaps of 1.08 eV and 1.16 eV were found for 35-aPNR and 35-zPNR respectively, which are close to the DFT calculated band gap of monolayer phosphorene (0.95 eV). It should be noted that the band gaps calculated at DFT-GGA level are typically underestimated compared to the experimental values or GW or hybrid functionals calculations. However, the focus of this study is not to quantitatively obtain band gaps of PNRs but to reveal the general trend and underlying physics of band gap modulation of PNRs by the quantum confinement and external electric field. Therefore, the underestimate of band gaps do not affect our main conclusions. The theoretical back ground of the transport and current density calculations were described in the APPENDIX of this article.

\section{Geometric structures}
The calculated lattice parameters of monolayer phosphorene lattice are $a_{1}=3.33$ {\AA} and $a_{2}=4.63$ {\AA}, respectively, which are in agreement with results of previous calculation.\cite{Liu2014AN,Qiao2014NC,Cai2014SR} Hydrogen saturated zigzag (zPNRs) and armchair (aPNRs) PNRs were constructed from the optimized phosphorene lattice as shown in \textbf{Fig. 1}. The number of zigzag lines (dimmer lines) across the zPNRs (aPNRs), $N$, is used to indicate the width of a PNR. Up on structural relaxation, the interior part of the nanoribbons experiences negligible structural changes and the P-P bond length decreased from 2.28 \AA [ref. 17] in bulk phosphorene to 2.24 \AA, while the edges of the zPNRs show some degree of deformation (see \textbf{Fig. 1}). The bonding angle in the edge of zPNRs is 98.7 degree compared to 103.8 degree in the central region. This is different from the case of bare zPNRs, which have larger bonding angles in the edge because of the edge reconstruction.\cite{Peng2014MRE,Guo2014JPCC}

\begin{figure}[ptb]
\centering
\includegraphics [width=0.75\textwidth]{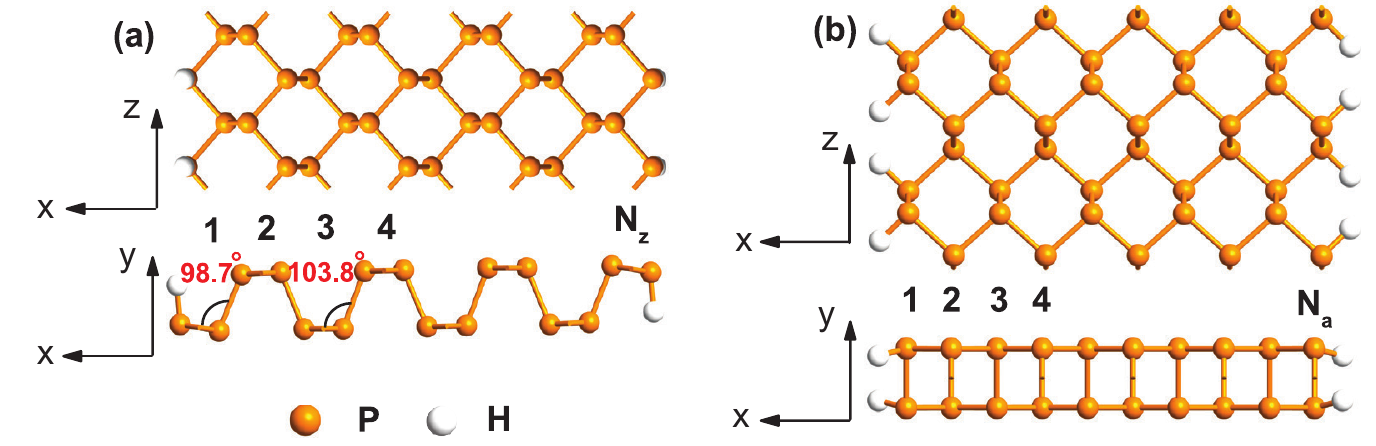}\newline
\caption{Top view and side view of the optimized geometry structure of hydrogen saturated phosphorene nanoribbons (PNRs): (a) zigzag phosphorene nanoribbons (ZPNRs) and (b) armchair phosphorene nanoribbons (APNRs).}%
\label{fig1}%
\end{figure}

\section{Energy band structures}

Different from hydrogen saturated graphene and MoS$_{2}$ nanoribbons which exhibit chirality-dependent electronic properties, {\em i.e.}, armchiar nanoribbns are semiconducting but zigzag nanoribbons are metallic, both armchair and zigzag hydrogen saturated PNRs inherit property of 2D phosphorene and exhibit semiconducting characteristics. The calculated band structures of an aPNR of width 10 (10-aPNR) and a zPNR of width 8 (8-zPNR) are presented in \textbf{Fig. 2}. At the level of the Perdew-Burke-Ernzerhof (PBE) functional, 10-aPNR has a direct band gap of 1.17 eV located at the $\Gamma$ point in the reciprocal space, while 8-zPNR possesses a band gap of 1.73 eV near the $\Gamma$ point which is almost direct. For both types PNRs, the calculated partial charge densities indicate that both VBM and CBM are contributed by hybridized $s$-$p$ states of the P atoms in the \emph{central} region of the nanoribbons. Therefore, hydrogen saturation provides perfect passivation of dangling bonds at the edges of PNRs.

\begin{figure}[ptb]
\centering
\includegraphics [width=0.9\textwidth]{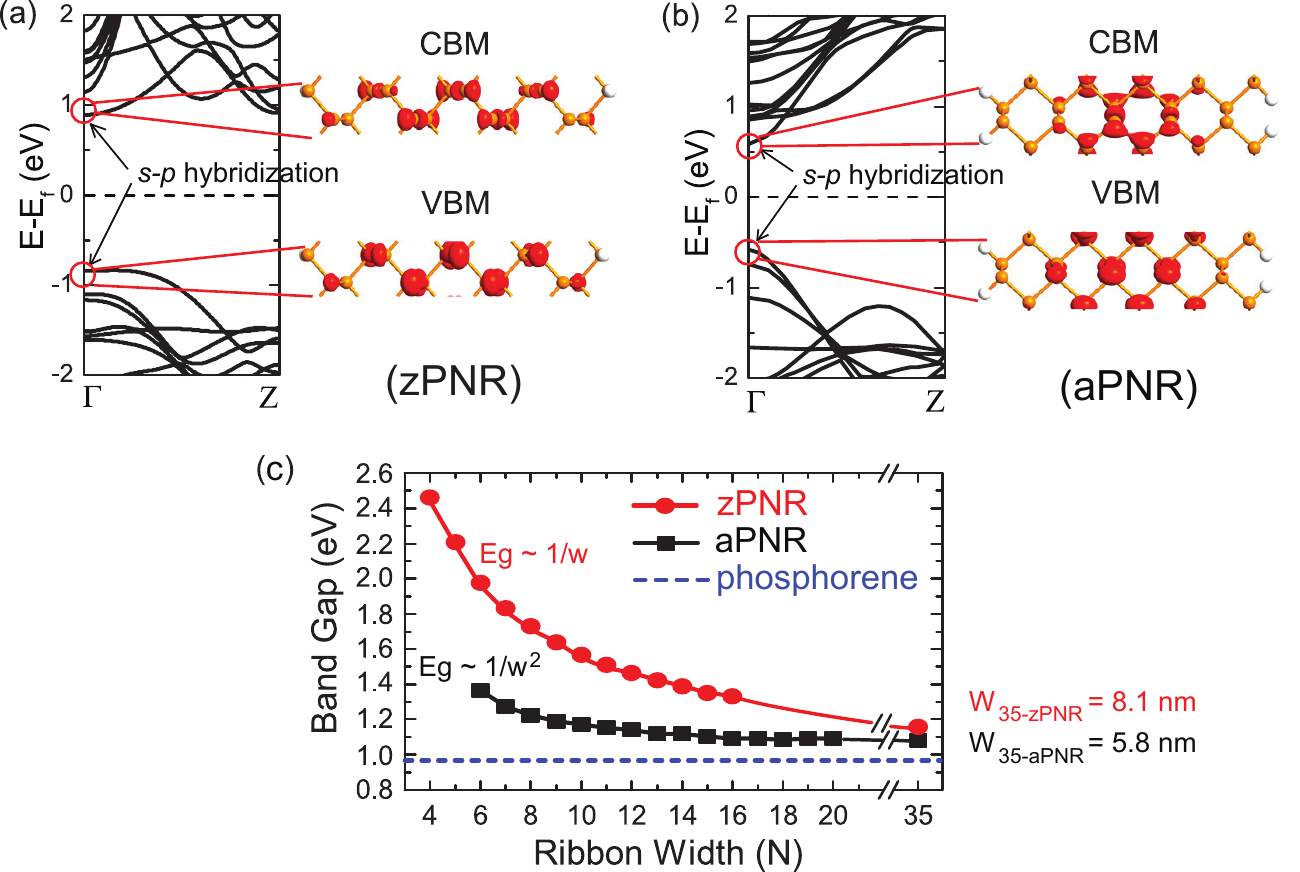}\newline
\caption{(Color online) Band structures and partial charge densities of the CBM and VBM of the (a) 8-zPNR and (b) 10-aPNR. The distribution of charge densities shows both electrons and holes are distributed in the center of ribbons. (c) Variation of band gaps of aPNRs (up to 5.8 nm) and zPNRs (up to 8.1 nm) as a function of ribbon width N. The scale law of band gap is $\sim 1/w$ for zPNRs and  $\sim 1/w^2$ for aPNRs, where $w$ is width of ribbon (in unit of Angstrom). The blue dashed line indicates the band gap of monolayer phosphorene.}%
\end{figure}

\begin{figure}[ptb]
\centering
\includegraphics [width=0.8\textwidth]{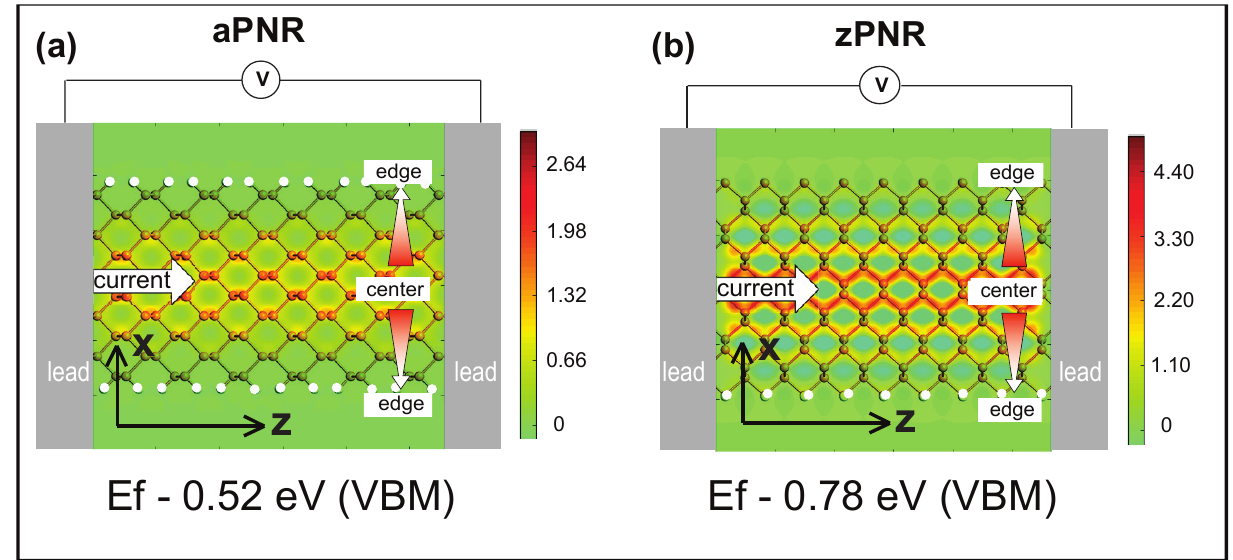}\newline
\caption{(Color online) The current density of aPNR (a) and zPNR (b) at the energy level sampling on the VBM, which shows the transport channels are at the center of ribbons.}%
\label{fig3}%
\end{figure}

We further examined dependence of band gaps of hydrogen saturated PNRs on ribbon width $N$. As shown in \textbf{Fig. 2(c)}, band gaps of both aPNRs and zPNRs decrease monotonously with increasing ribbon width and eventually converge to the band gap of 2D phosphorene. It is noted that the band gap of aPNRs approaches to the 2D limit faster than zPNRs. This
is because charges in aPNS are more localized in the central part of the nanoribbon, as shown by the charge density and isovalue bar in \textbf{Figs. 2(a) and 2(b)}, and therefore less affected by the edges, compared to zPNRs. In contrast, even though the band gaps of zigzag nanoribbons of MoS$_2$, BN and graphene vary monotonously with ribbon width, the band gaps of \emph{armchair} nanoribbons of MoS$_2$, BN and graphene nanoribbons all show an oscillatory behavior as a function of ribbon width, due to changes in the symmetry
of wave function of the edge states as the width of the ribbon increases. Compared to the chirality dependent properties of MoS$_{2}$ and graphene nanoribbons, the robust semiconducting behavior and monotonous width-dependence of band gap of PNRs make PNRs a more promising candidate than MoS$_{2}$ and graphene nanoribbons for PNRs-based FETs.

\section{Transport properties}

\subsection{Transport channels}

To study the transport properties of phosphorene nanoribbons, we first calculate the current density of both armchair and zigzag PNRs without defects at the energy level sampling on the valance band maximum (VBM), which can be approximately referred as applying a small bias. The dangling bonds are saturated by pseudo-hydrogen. The contour plots of current density of PNRs are shown in \textbf{Fig. 3}. Surprisedly, in contrast to other nanoribbons, such as graphene and MoS$_{2}$, the transport channels in both armchair \textbf{(Fig. 3(a))} and zigzag PNRs \textbf{(Fig. 3(b))} are in the interior of PNRs, and the electron densities decay from the center to two edges. The current flow direction is from left to right and the current density difference is demonstrated by warm color in the figures. Besides the current density, the local electron transmission pathway are shown in \textbf{Figs. 4(a) and 4(b)}. It is found that the armchair phosphorene has a characterize of inter-layer transport, whereas the zigzag phosphorene shows a characterize of intra-layer transport, which may be one of the reasons of anisotropic conductivity of phosphorene in the experiment.\cite{Liu2014AN}

\begin{figure}[ptb]
\centering
\includegraphics [width=0.8\textwidth]{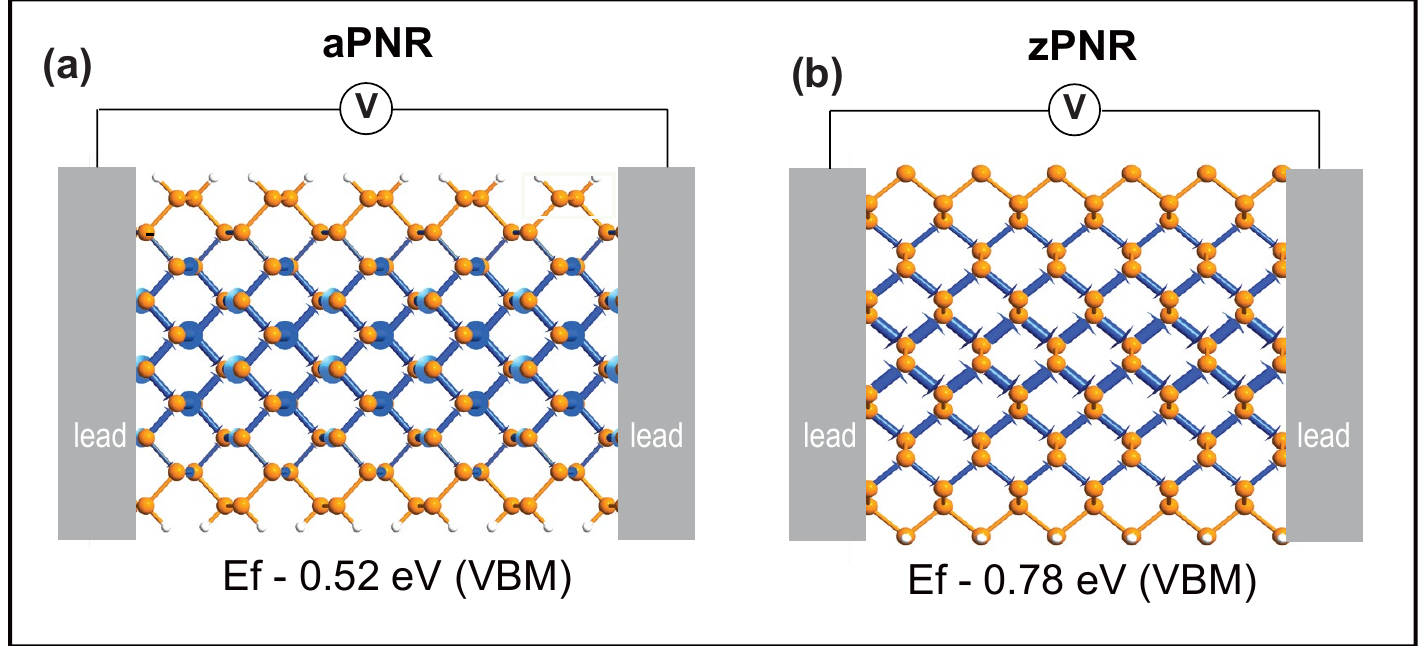}\newline
\caption{(Color online) The local electron transmission pathway of armchair PNRs (a) and zigzag PNRs (b) with the energy level sampling on the valance band maximum. }%
\label{fig4}%
\end{figure}

In order to understand the central transport behavior in phosphorene nanoribbons, we further calculate the charge density of both armchair and zigzag PNRs (\textbf{Fig. 2}). For both types of PNRs, the calculated partial charge densities indicate that both VBM and conduction band maximum (CBM) are contributed by hybridized $s$-$p$ states of the P atoms in the \emph{central} region of the nanoribbons. As a matter of fact, the H-P bond is stronger than the P-P bond such that the edge states of PNRs are located deep in the bands. This is fundamentally different from armchair graphene and MoS$_{2}$ nanoribbons whose VBM and CBM consist of mainly edge states. Therefore, the electronic properties of armchair graphene and MoS$_{2}$ nanoribbons depend strongly on the edge symmetry of the nanoribbons.\cite{Dolui2012AN,Son2006PRL,Yue2012JPCM} Having the transport channels in the central region of PNRs means that carrier transport under a low bias (for electrons in CBM or holes in VBM) is robust against edge disorder or defects, a desired property for device applications.

\subsection{Defect effect}

It is well known that the transport channels in most nanoribbons are located at the two edges.\cite{Kim2008NN,Zheng2008PRB,Yue2012JPCM} A small amount of edge-defects, such as vacancies and impurities, can strongly suppress the carrier mobility and device conductivity as Coulomb blockage and scattering centers,\cite{Sols2007PRL,Shen2012PRB} which hinders the development of nanoribbon devices. The unique central transport channels in PNRs may render the electrons travelling in PNR-based devices insensitive to scattering of edge vacancies and impurities. To verify this hypothesis, we calculate the current density at VBM of zPNRs with an edge P vacancy and $H_2$-impurity, which are shown in \textbf{Figs. 5(a) and 5(b)}. As can be seen, the current really immune to different types of edge defects.

\begin{figure}[ptb]
\centering
\includegraphics [width=0.8\textwidth]{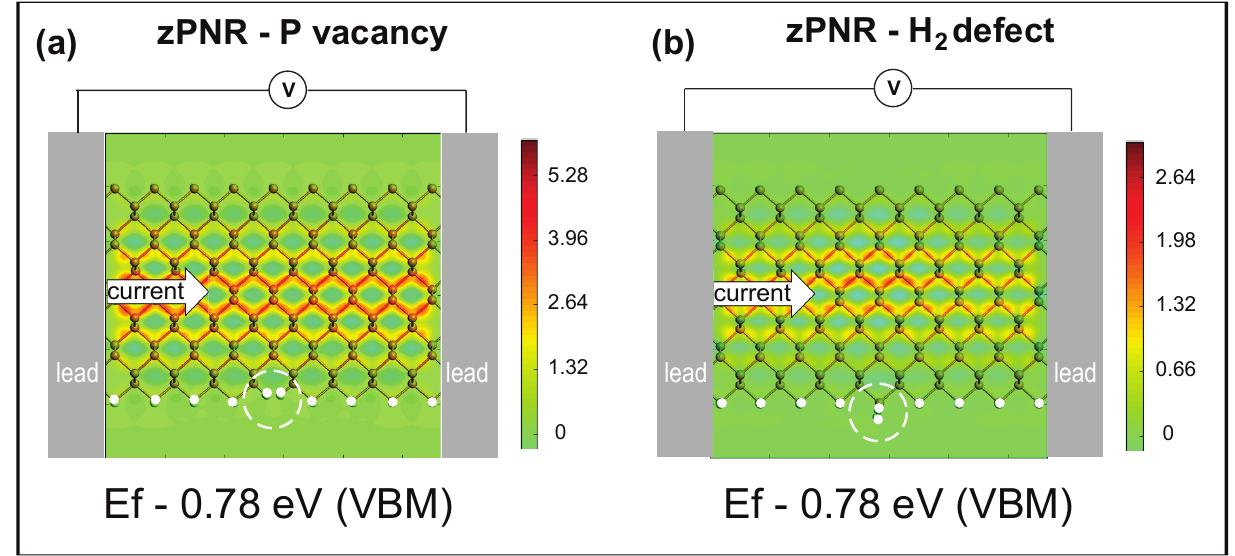}\newline
\caption{(Color online) The current density of zPNR with a P vacancy defect (a) and H2 defect (b) at the energy level sampling on the VBM.}%
\label{fig5}%
\end{figure}

\subsection{Bias voltage effect}

If more carriers are explicitly injected into the system, for example applying a high bias, they may want to populate deep bands instead of the VBM in order to minimize the electrostatic interaction between them. Thus, under a high bias, the edge transport channels might be opened since these deep bands are contributed by the orbitals of edge atoms. To demonstrate this assumption, i.e., whether edges conduct under a high bias, we first plot the weight of states of outermost edge P atoms of the aPNR onto the band structure [\textbf{Fig. 6(a)}] to find the ``edge" bands. Because the zPNR has the similar deep edge bands, the case of zPNR is not discussed here. As can be seen, the energy band at -1.02 eV is mainly contributed by the orbitals of outermost edge P atoms. We next calculate the current density from left to right going states at this special deep energy level (-1.02 eV) as shown in \textbf{Fig. 6(d)}. As can be seen, both central and edge transport channels are opened if high bias voltage cover deep bands. Although the transmission spectrum is different from the energy band,\cite{Das2014NL} we can estimate the trend based on an aPNR Current-Voltage (I-V) curve (\textbf{Fig. 6(e)}). The trend is (I) under a very low bias (Region I), the semiconducting PNRs are not electrically conducting (\textbf{Fig. 6(b)}); (II) under a small bias (Region II), the PNRs become conducting and the conductive channels are in the center of ribbons (\textbf{Fig. 6(c)}); (III) under a high bias (Region III), both the central and edge channels are conducting (\textbf{Fig. 6(d)}). In this paper, we mainly focus on the case of low source-drain bias because the sournce-drain bias is usually only a few voltage in the phosphorene experiments.\cite{Li2014NN,Liu2014AN}
\begin{figure}[ptb]
\centering
\includegraphics [width=0.9\textwidth]{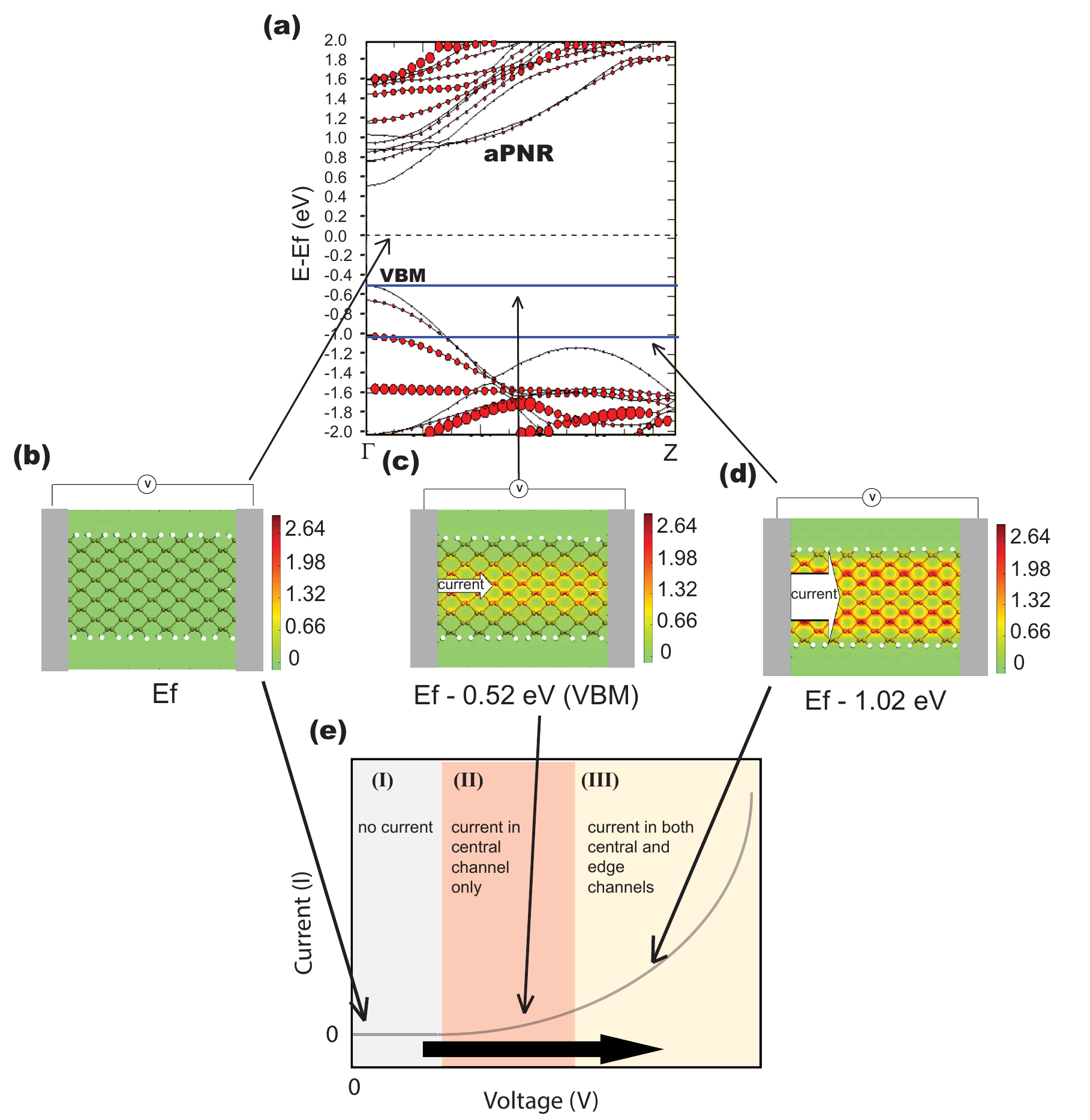}\newline
\caption{(Color online) (a) Band structure of the armchair PNR. The red solid circles indicate the weight of states of outermost edge P atoms. (b)-(d) The current density at different energy levels (Ef, VBM and Ef-1.02 eV). (e) A schematic diagram of the aPNR I-V curve.}%
\label{fig6}%
\end{figure}

\section{Electric field effect}

\begin{figure}[ptb]
\centering
\includegraphics [width=0.55\textwidth]{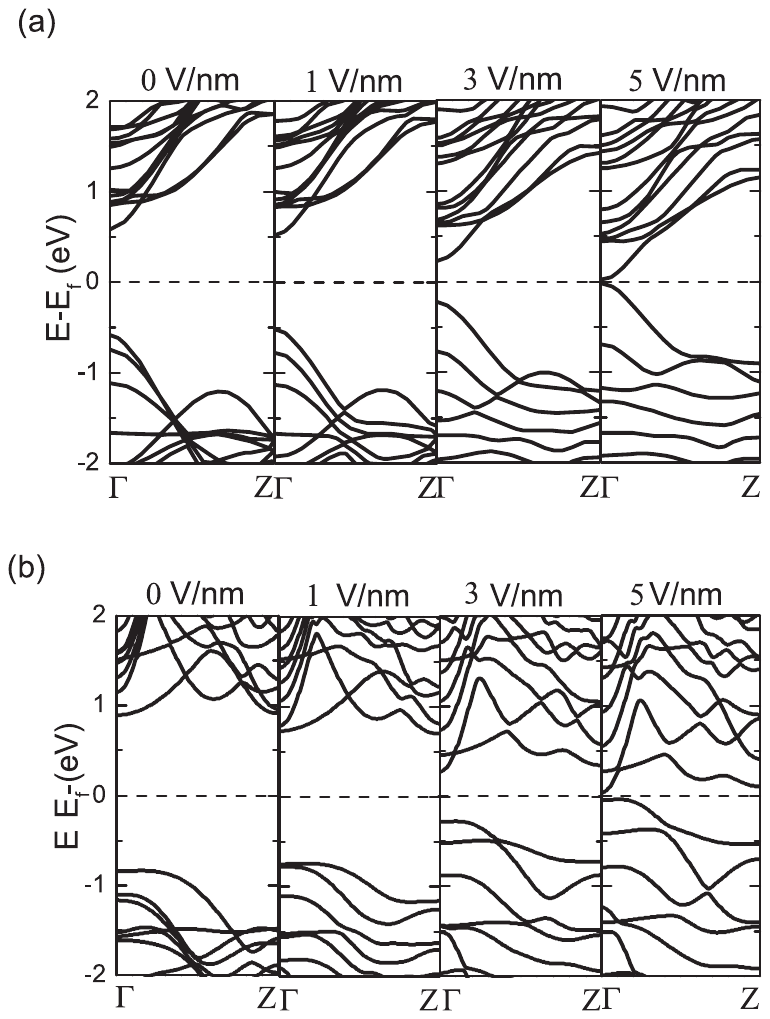}\newline
\caption{(Color online) Band structures of the 10-aPNR (a) and 8-zPNR (b) under electric fields. The figures show VBM and CBM energy levels shift and split both in aPNRs and zPNRs, leading to a reduction in the band gap and closure at the Fermi level.}%
\label{fig7}%
\end{figure}

\begin{figure}[ptb]
\centering
\includegraphics [width=0.90\textwidth]{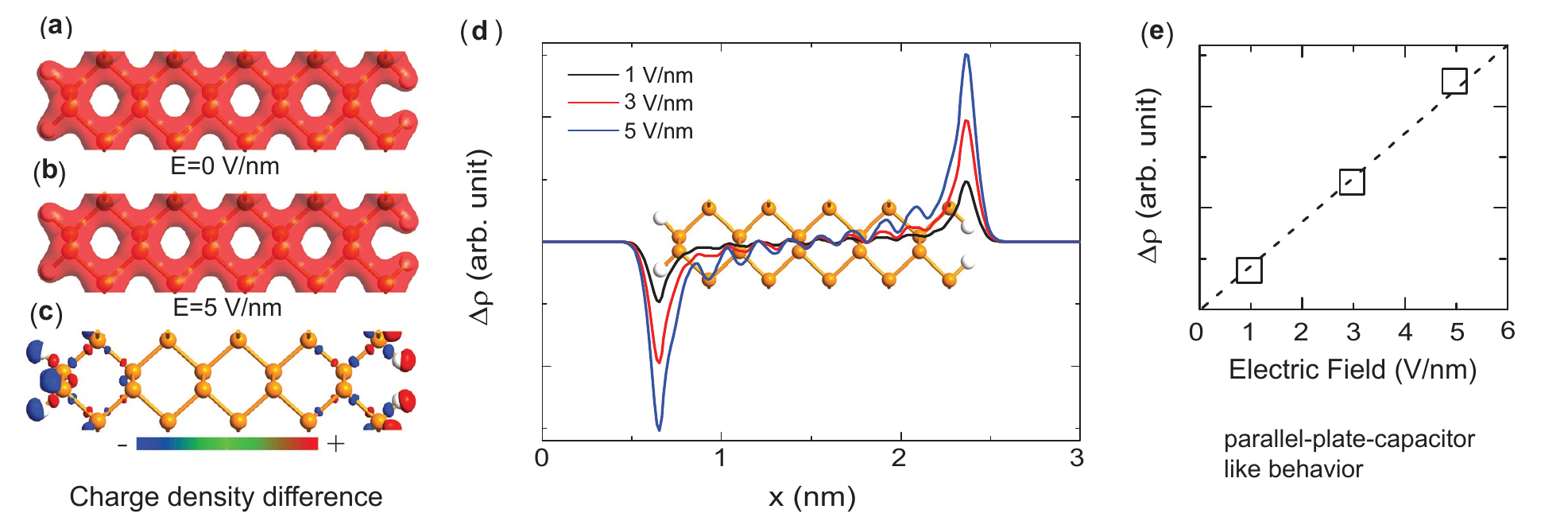}\newline
\caption{(Color online) (a) and (b) The total charge density of 10-aPNRs under a 0~V/nm and 5~V/nm electric field. (c) The charge density difference of (a) and (b). (d) Electric field induced charge density difference $\Delta\rho=\rho_{E_{ext}}-\rho_{0}$ as a function of the position across the nanoribbon of the 10-aPNR. $\Delta\rho$ is averaged over the plane perpendicular to the electric field direction (yz-plane). (e) the accumulated charge $\Delta\rho_{acc}$ as a function of the external electric field, which shows a parallel-plate-capacitor-like behavior.}%
\label{fig8}%
\end{figure}

 For the field-effect-transistor application, large on/off ratio is required, which means the electronic structure should be sensitive to the gate voltage or external electric field as large as possible.\cite{Koenig2014APL} We, thus, investigate the band modulation of PNRs by an external electric field. We first considered an external electric fields applied perpendicular to the plane of a PNR, but found it has no effect to the band gap of the PNR. We thus conclude that a planar phosphorene nanostructure with a longitudinal gate will not be electronically responsive. On the other hand, when an in-plane transverse electric field is applied across the nanoribbon, significant changes in band structures are induced for both aPNRs and zPNRs. \textbf{Figure 7} shows the external electric field-dependence of the band gaps of aPNR and zPNR. Compared to the band structure without electric field (0 V/nm), the energy bands of degenerate edge states above CBM and below VBM under the electric field show certain degree of splitting and localization [see very right panel in \textbf{Fig. 7(a) and 7(b)}], which pushes the CBM and VBM closer to the Fermi level, leading to the band gap narrowing. Note that this behavior is in contrast to the electric field effect on band structures of other nanoribbons, in which the CBM and VBM are consisted of edge states. Under an external electric field, these CBM and VBM bands are split, narrowing the band gap. \textbf{Figures 8(a) and 8(b)} show the total charge density distribution of 10-aPNRs without electric field and under a 5~V/nm field. \textbf{Figure 8(c)} shows the charge density difference of \textbf{Figures 8(a) and 8(b)}. As can be seen, the in-plane transverse electric field induces obvious charge redistributions, with holes in the VBM shifting in the direction of the field, towards the edge with low electrical potential, and electrons in CBM in opposite direction. What we see here is actually a giant Stark effect (GSE).\cite{Zhang2008PRB,Zheng2008PRB} The applied transverse electric field breaks the symmetry of the nanoribbon, induces a difference of electrostatic potential across the nanoribbon, splits the edge energy levels above CBM and below VBM, and leads to the band gap narrowing. This can be further verified by a quantitative charge density redistribution analysis. The electric field induced charge density difference, $\Delta\rho$, as a function of the position across the nanoribbon is shown in \textbf{Fig. 8(d)}. $\Delta\rho$ is defined as $\Delta\rho=\rho_{E_{ext}}-\rho_{0}$, where $\rho_{E_{ext}}$ and $\rho_{0}$ are the charge densities with and without applied external electric fields, respectively. $\Delta\rho$ has also been averaged over the plane perpendicular to the electric field direction (xy-plane) for easy understanding. As can be seen, charges accumulate at the positive potential ribbon edge, while deplete at the negative potential ribbon edge. When the electric field increases, there are more charge accumulation and depletion at each edge of the ribbon, which will further narrow the band gap and finally close it. The accumulated charge $\Delta\rho_{acc}$, which is defined as the integration of $\Delta\rho$ from the ribbon middle point to the ribbon edge ($\Delta\rho_{acc}=\int\nolimits_{x_{edge}}^{x_{middle}}\Delta\rho (x)dx$), as a function of applied external electric fields is also shown in \textbf{Fig. 8(e)}. There is an obvious linear relationship between $\Delta\rho_{acc}$ and the field. This parallel-plate-capacitor-like behavior directly indicates the Stark effect.

\begin{figure}[ptb]
\centering
\includegraphics [width=0.95\textwidth]{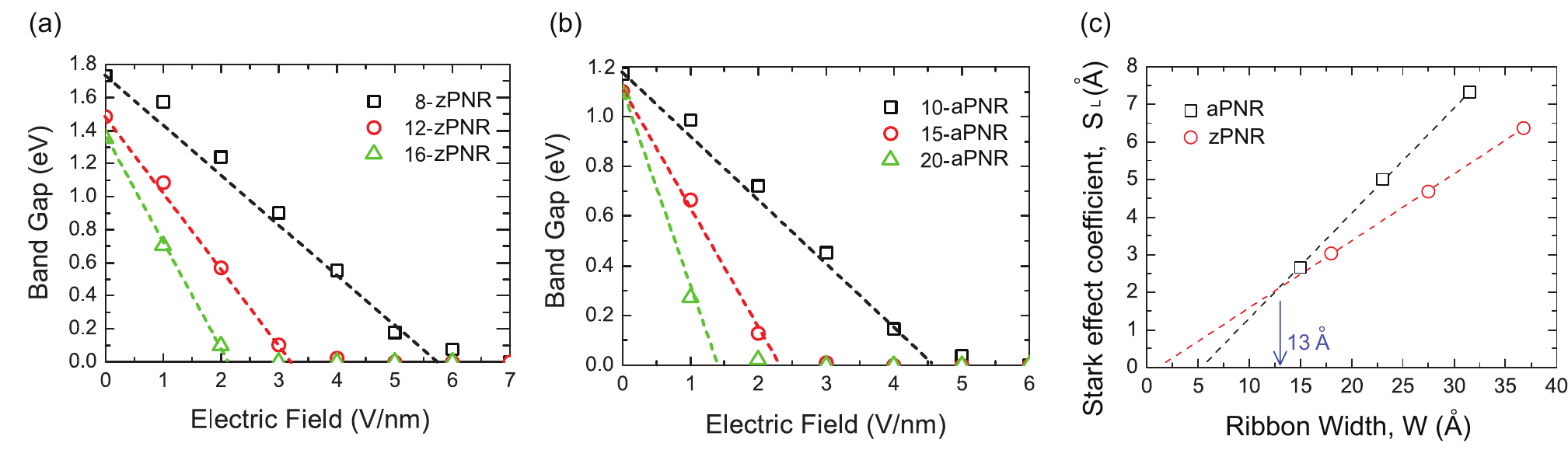}\newline
\caption{(Color online) Variation of band gaps of zPNRs (a) and aPNRs (b) as a function of external electric field. Three different widths of ribbons were considered for each case. (c) Calculated giant Stark effect coefficient S$_{L}$ as a function of the ribbon width W.}%
\label{fig9}%
\end{figure}

 To estimate the intensity of the GSE, different widths ($N=8, 12, 16$ for zPNR and $N=10,15,20$ for aPNR) were considered in each case in \textbf{Fig. 9(a) and 9(b)}. Overall, the band gap decreases linearly with increasing electric field, similar to the trend found in the cases of armchair nanoribbons of MoS$_{2}$ and BN .\cite{Dolui2012AN,Park2008Nl,Zhang2008PRB,Zheng2008PRB} When the field reaches a certain critical value, the band gap becomes zero, exhibiting a field-induced metal-insulator-transition. For both aPNRs and zPNRs, the electric field effect is more appreciable in wider nanoribbons. For example, based on PBE, an electric field as high as 6 V/nm is required to close the gap of 8-zPNR, compared to the critical field of 2 V/nm for 16-zPNR. The enhanced sensitivity to electric field in wider PNRs is important, as it means the band gap of a wide PNR can be tuned by a relatively weak electric field. The method is applicable to experimentally available nanoribbons which are typically of more than tens of nanometers in width. Next, we took the linear part of the band gap curves in \textbf{Fig. 9(a) and 9(b)}, then calculated the linear GSE coefficient $S_{L}$ using $\frac{\Delta E_{g}}{\Delta E_{ext}}=-eS_{L}$, where $E_{ext}$ is the external electric field and $e$ is the electron charge. The external electric field induces a potential of $eE_{ext}x$ across the ribbon, therefore, the band gap change is approximately $\Delta E_{g}=eE_{ext}\left(  \left\langle x\right\rangle _{cb}-\left\langle x\right\rangle _{vb}\right)$, where $\left\langle x\right\rangle _{cb}$ and $\left\langle x\right\rangle_{vb}$ are the centers of the CBM and VBM respectively.\cite{Zheng2008PRB} Since $\left(  \left\langle x\right\rangle _{cb}-\left\langle x\right\rangle _{vb}\right)  $ is proportional to the ribbon width $W$ (in unit of Angstrom), using the two equations above we can get the linear scaling law of the GSE coefficient $S_{L}$ on the ribbon width $W$, $S_{L}=\alpha W+C$, where $\alpha$ is the slope of the line and $C$ is a constant. Our calculated GSE coefficient $S_{L}$ as a function of ribbon width $W$ is given in \textbf{Fig. 9(c)} and it demonstrates the linear relationship of $S_{L}$ and $W$, following the GSE mechanism. The slopes of aPNR and zPNR are 0.27 and 0.17, respectively, which are much higher than that of CNTs \cite{Son2005PRL} because of the reduced screening of the electric field. As can be seen, the two lines cross at about 13 \AA. Since the GSE coefficient $S_{L}$ indicates the ability of band gap tuning by electric fields, we know that the electronic structure of wider aPNRs (ribbon widths $>$13 \AA) is more sensitive to the external electric field than zPNRs.

\section{Device design}

This giant Stark effect can be utilized to design PNRs-based transistors. Taking zigzag PNRs in \textbf{Fig. 1(a)} as an example, we demonstrate that the PNR-FETs can have high on/off ratio. The designed FET is shown in \textbf{Fig. 10(a)}, where unsaturated zPNRs are used as metallic electrodes.\cite{Guo2014JPCC,Peng2014MRE} In the middle part of the device, saturated zPNRs (by pseudo-hydrogen) serve as tunneling barriers with top and bottom gates to generate a transverse electric field. This is an all-phosphorus based FET which can avoid the metal-semiconductor interfacial contact effect on the transport property. The calculated transmission spectrum of the zPNR based FET under zero and 7 V/nm electric fields without source-drain voltage are shown in \textbf{Fig. 10(b)}. Due to the semiconducting characteristic of saturated zPNRs, there is no transmission states near the Fermi level under zero electric field with a transmission gap of 1.9 eV. When an electric field of 7 V/nm is applied, a transmission peak emerges at the Fermi level with sufficient large dispersion (-0.1 eV to +0.2 eV), not a usual Van Hove-like singularity in one-dimensional materials. This means that the on-state of the FET can be stable at room temperature. The transmission eigenchannels at E$_{f}$ and at the (0,0) point of the $k$-space, presented in \textbf{Fig. 10(c)}, vividly illustrate off- and on-state of the zPNRs based FET controlled by a dual-gate induced electric field. Without an external transverse electric field, the calculated transmission eigenvalue is 0.001 G$_0$ (G$_0$ = 2$e^2/h$, where $e$ and $h$ are the electron charge and Planck`s constant, respectively), and thus the transmission channels are blocked, resulting an off-state. On the contrary, with the external transverse field of 7 V/nm, the transmission eigenvalue reaches the value of 1 G$_0$ and the transmission channels are opened (on-state), with on/off ratio of 10$^3$. Because of the quantum confinement effect, the on/off ratio of PNR-FETs is 2 order lower than phosphorene FETs (10$^5$) [ref. 21], but comparable to graphene and MoS$_2$ nanoribbon FETs .\cite{Koenig2014APL,Li2014NN} The calculated density of states (DOS) of the scattering region under an external electric field of 7 V/nm is shown in the inset of \textbf{Fig. 10(b)}. Our calculated DOS shows a peak at the Fermi level, which implies a strong correlation between transmission and DOS. The physics of such strong correlation is that the transport at the Fermi level is dominated by resonant tunneling through interface states (see \textbf{Fig. 10c}), not barrier tunneling.

\begin{figure}[ptb]
\centering
\includegraphics [width=0.85\textwidth]{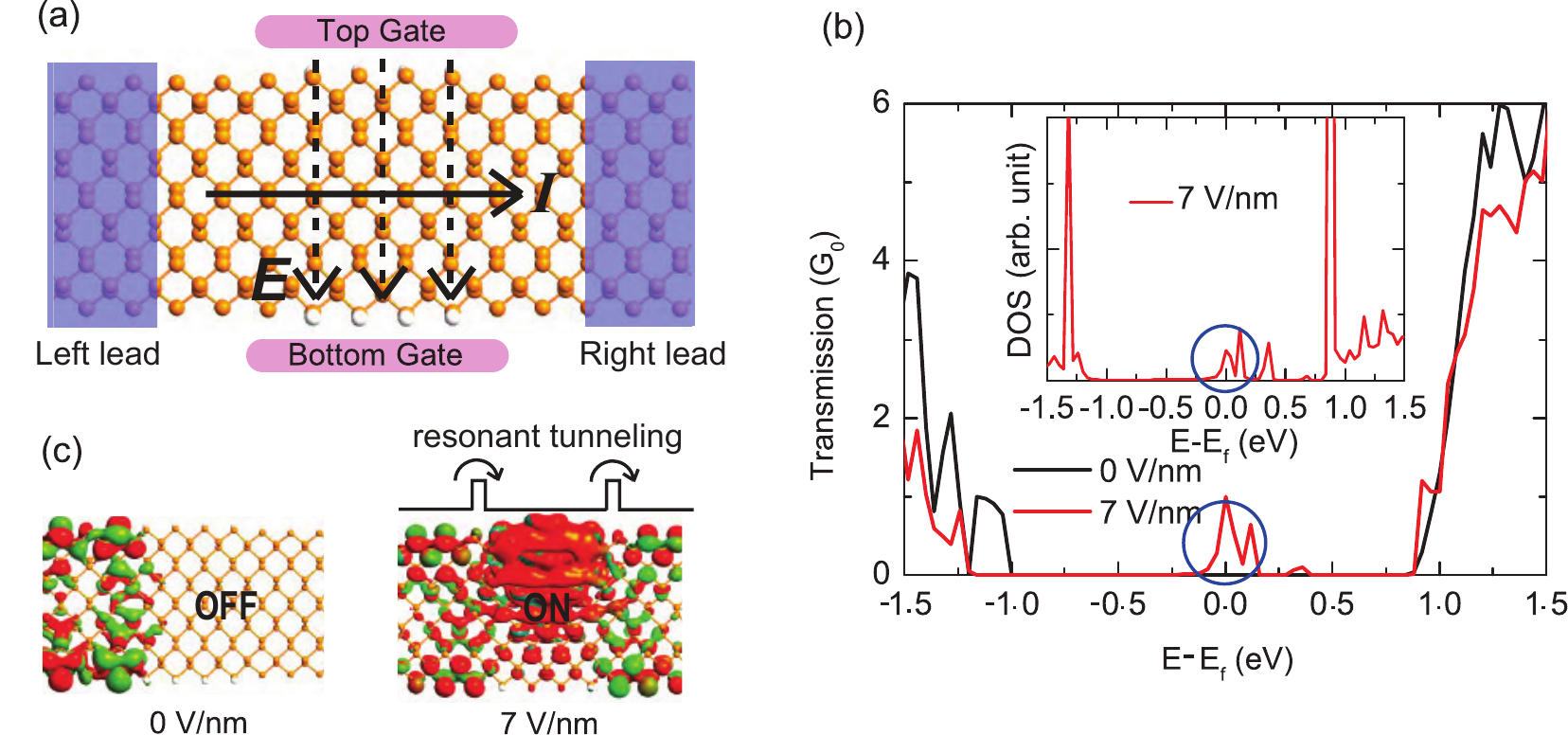}\newline
\caption{(Color online) (a) Top view of a dual-gate field effect transistor based on zPNR. Semi-infinite metallic bare zPNRs serves as two leads while hydrogen saturated zPNR is used as semiconducting channel (scattering area). (b) Transmission spectrum under E = 0 V/nm (black line) and E = 7 V/nm (red line). Inset: the DOS of the hydrogen saturated zPNR (the scattering region) under an external electric field of 7 V/nm. (c) Transmission eigenstates at E$_{f}$ and at the (0,0) point of the k space under E = 0 V/nm and E = 7 V/nm. }%
\label{fig10}%
\end{figure}

\section{Conclusions and discussions}

In conclusion, based on both electronic and transport calculations, we report unique electronic band structures and carrier transport properties in phosphene nanoribbons. For example, 1) the central atoms contribute to the states of VBM and CBM; 2) the transport is achiral and robust under low  bias, which can offer more feasibility of using PNRs to fabricate nano-scale FETs easily. Meanwhile, the electric field can effectively switch on transport channels of semiconducting PNRs due to the giant Stark effect. Thus, high on/off ratio can be demonstrated in a dual-gate PNR-FET. Furthermore, our calculation results imply that the direct bandgap behavior of aPNRs is not affected by the electric field during bandgap modulation, which indicates the potential application of aPNRs in opto-electronics as well. Furthermore, recently zPNRs are reported having magnetic behavior which sheds light on PNRs in spintronics applications.\cite{Du2015SR,Zhu2014APL}

\section{Acknowledgement}

The Authors thank M. G. Zeng for helpful discussion on the FET devices. The first-principles calculations were carried out on the GRC-NUS high-performance computing facilities.

\section{Appendix}

In the transport calculations, the device system is compose of a finite
central (C) region sandwiched between semi-infinite left (L) and right (R)
electrodes. To decompose the Kohn-Sham equation into these three regions, the
local orbital basis set $\left\{  {\psi}_{i}\right\}  $ is adopted. Under such
a condition, the Hamiltonian $H$ can be written as%

\[
H_{ij}=\left\langle {\psi}_{i}\right\vert {-\frac{\hbar^{2}}{2m}\nabla
^{2}+V_{eff}}\left\vert {{\psi}_{j}}\right\rangle ,
\]
where ${V_{eff}}$ is the Kohn-Sham effective potential, and overlap matrix $S$
is defined as
\[
S_{ij}=\left\langle {\psi}_{i}\right\vert \left.  {{\psi}_{j}}\right\rangle .
\]

Thus, the Green's function for the central region to describe the device can
be evaluated by%
\[
G(E)=\left[  (E+i\delta_{+})S-H-\Sigma_{L}(E)-\Sigma_{R}(E)\right]  ^{-1},
\]
where $S$ and $H$ are the overlap matrix and the system Hamiltonian of the
central region, respecitvely, $\Sigma_{L/R}$ is the self energy which
describes the coupling between the central region and the left/right
electrode, and $\delta_{+}$\ is a positive infinitesimal.

Then, the density matrix $D$ can be calculated from the above Green's function%

\[
D=-\frac{1}{\pi}{\text{Im}}\int G(E)f(E-\mu)dE,
\]
where, $f$ and $\mu$ are the Fermi function and the chemical potential
respectively. After that one gets the electron density%

\[
\rho(\mathbf{r})=\sum_{i,j}{\psi}_{i}(\mathbf{r})D_{i,j}{\psi}_{{j}%
}(\mathbf{r}).
\]

According to the Kohn-Sham theory, the Hamiltonian $H$ depends on the electron
density $\rho(\mathbf{r})$. Therefore, one can use a self-consistent iteration
scheme to find the ground state of the system. When the system finally
converges to its ground state, the transmission can be calculated using
\[
T(E)=Tr[\Gamma_{L}(E)G(E)\Gamma_{R}(E)G^{\dagger}(E)],
\]
where $\Gamma_{L/R}$ is the coupling matrix for the left/right electrode,
$G/G^{\dagger}$ is the retarded/advanced Green's function matrix. The current
density is evaluated by%

\[
J(r,E)=-\frac{e\hbar}{4\pi m}\int\sum_{i,j}G_{i,j}^{<}(E)\psi_{i}\nabla
\psi_{j}dE,
\]
where $G^{<}(E)=G(E)\Gamma_{L/R}(E)G^{\dagger}(E)w(E)$ is the lesser Green
function, in which, $G$ is the retarded Green function, $\Gamma$ is the
coupling matrix, and $w(E)=f_{R}(E)-f_{L}(E)$ is a spectral weight given by
the left/right Fermi function. The positive current means current from left to
right (warm color in the contour plot). Through the definition of the spectral
weight $w(E)$, one can use the current density to analyze zero bias calculations.

%

\end{document}